\documentclass[showpacs,aip,jap,reprint,amsmath,amssymb]{revtex4-1}
\usepackage{graphicx}

\begin{document}
\title{Structural and magnetic properties of Co-Mn-Sb thin films}
\author{Markus Meinert}
\email{meinert@physik.uni-bielefeld.de}
\author{Jan-Michael Schmalhorst}
\email{jschmalh@physik.uni-bielefeld.de}
\author{Daniel Ebke}
\author{Ning-Ning Liu}
\author{Andy Thomas}
\author{G\"unter Reiss}
\affiliation{Thin Films and Physics of Nanostructures, Department of Physics, Bielefeld University, 33501 Bielefeld, Germany}
\author{Jaroslaw Kanak}
\author{Tomasz Stobiecki}
\affiliation{Department of Electronics, University of Science and Technology, 30059 Krakow, Poland}
\author{Elke Arenholz}
\affiliation{Lawrence Berkeley National Laboratory, Berkeley, CA 94720, USA}
\date{\today}

\begin{abstract}

Thin Co-Mn-Sb films of different compositions were investigated and utilized as electrodes in alumina based magnetic tunnel junctions with CoFe counter electrode. The preparation conditions were optimized with respect to magnetic and structural properties. The Co-Mn-Sb/Al-O interface was analyzed by X-ray absorption spectroscopy and magnetic circular dichroism with particular focus on the element-specific magnetic moments. Co-Mn-Sb crystallizes in different complex cubic structures depending on its composition. The magnetic moments of Co and Mn are ferromagnetically coupled in all cases. A tunnel magneto resistance ratio of up to 24\% at 13\,K was found and indicates that Co-Mn-Sb is not a ferromagnetic half-metal. These results are compared to recent works on the structure and predictions of the electronic properties.
\end{abstract}

\maketitle

\section{Introduction}\label{intro}
Half-metallic materials are characterized by a gap in the density of states for minority (or majority) electrons at the Fermi energy, giving a spin polarization of $P=100$\%. Therefore these materials are of great interest for spintronic devices such as magnetic sensors, nonvolatile memories\cite{tre01} or programmable logic devices\cite{rei05b,apl2006V89S012502}. Taking  Julli\`ere's simple model \cite{jul75} of the tunnel magnetoresistance, an infinite TMR ratio is expected if two half-metals are used as ferromagnetic electrodes in magnetic tunnel junctions (MTJs). 

Several half and full Heusler compounds have been predicted to be half metals\cite{bus83,asa98,pap02,mor05,ded05} and implemented in MTJs. The largest spin polarizations in Al-O based junctions were reported for the full Heusler compounds Co$_2$Cr$_{1-x}$Fe$_x$Al\cite{hir04} and  Co$_2$YZ (Y = Mn or Fe; Z = Si or Al)\cite{hue05,miy06,rei06,kub06,are07a,fel08} after Tanaka {\it et al.}\cite{moo99a} started investigating NiMnSb in 1999. Co$_2$MnSi electrodes combined with MgO barriers yield even higher TMR ratios\cite{yam08}. However, the highest TMR ratios are currently found in Co-Fe-B/MgO/Co-Fe-B junctions\cite{ohn08}.

Ideally, a full Heusler compound X$_2$YZ crystallizes in $L2_1$ structure whereas a half Heusler compound XYZ crystallizes in $C1_b$ structure. Real Heusler compounds often show a certain degree of atomic disorder and thus deviate from the ideal structures. Generally, this has a strong influence on the electronic properties of the films such as the spin polarization.

Here, Co-Mn-Sb is implemented as an electrode in magnetic tunnel junctions. CoMnSb was thought to be a half Heusler compound and the structural, magnetic and electronic properties of bulk CoMnSb were controversially discussed in the literature. Galanakis\cite{gal02} assumed the $C1_b$ structure for CoMnSb and calculated its band structure. The total spin magnetic moment was almost $3\,\mu_B$ per formula unit (f.u.), while the Mn ($3.18\,\mu_B$) and the Co ($-0.13\,\mu_B$) spin moments were antiferromagnetically aligned. Presuming a typical lattice constant of {5.9\,\AA{}} this corresponds to a low temperature saturation magnetization of 540\,kA/m.  Half metallicity  with a large spin-down gap of about 1\,eV was proposed for this structure. However, a total magnetic moment of $3\,\mu_B$ per f.u. is in contradiction to an experimentally observed value of about $4\,\mu_B$ per f.u.\cite{haa87}. As discussed by Tobola {\it et al.}\cite{pie00} this larger moment can be explained by a more complex structure (space group $Fd\bar3m$), which was originally proposed by Senateur {\it et al.}\cite{fru72} in 1972. A narrow spin-down gap is found for this structure, but the half-metallic character is preserved. The calculated mean Mn and Co magnetic moments are $3.51\,\mu_B$ and $0.48\,\mu_B$, respectively, and are expected to couple ferromagnetically. 

Kaczmarska {\it et al.}\cite{sko99} could also reproduce a magnetic moment of nearly $4\,\mu_B$ per f.u.\ by their calculations, where a `disordered' $L2_1$ structure was assumed. The two X sublattices were occupied by 50\% Co and 50\% vacant sites and the Y by Mn and Z by Sb. The ferromagnetically coupled Mn and Co magnetic moments were calculated to be $3.60\,\mu_B$ and $0.30\,\mu_B$ and half-metallicity disappeared. 

Recently, Ksenofontov {\it et al.}\cite{fel06} revised the structure of CoMnSb bulk samples using XRD, NMR and M\"ossbauer spectroscopy. They assigned it to space group $Fm\bar 3m$ and presented the CoMnSb structure as an alternation of Co$_2$MnSb and MnSb structural units with a calculated Co moment of $0.55\,\mu_B$ and a mean Mn moment of $3.48\,\mu_B$. Taking the experimental lattice parameter of 11.73\,\AA{} into account this corresponds to a magnetization of 740\,kA/m. CoMnSb crystallized in this structure is predicted to be not half-metallic. 

In this paper, we present investigations of Co-Mn-Sb thin films of different compositions implemented in MTJs with Al-O barrier and Co-Fe counter electrode. In Sec. \ref{SecIII} we analyze the bulk properties of the films, namely magnetization, interdiffusion with the buffer, element specific magnetic moments and crystal structure. Optimized growth conditions of the films were obtained with respect to these quantities. In Sec. \ref{SecIV} we discuss the properties of the Co-Mn-Sb / Al-O interface and the tunnel properties of the corresponding magnetic tunnel junctions.

\section{Experimental}\label{SecII}
The samples were deposited at room temperature by DC-magnetron sputtering in a sputter tool with a base pressure of $2\cdot 10^{-7}$\,mbar. Thermally oxidized Si(100) wafers and 200\,nm thick Si$_3$N$_4$ membranes with a window size of 3\,mm $\times$ 3\,mm were used as substrates. First, a vanadium buffer was deposited, then, the Co-Mn-Sb films were prepared from three different targets. 
\begin{table}[h]
\begin{tabular}{cccc}
{\bf target}&Co&Mn&Sb\\ \hline 
I & 29.0 & 32.1 & 38.9 \\ 
II & 33.3 & 33.3 & 33.3 \\ 
III & 35.1 & 29.4 & 35.5 \\ 
\end{tabular}
\hfill
\begin{tabular}{cccc}
{\bf film}&Co&Mn&Sb\\ \hline 
I & 32.4 & 33.7 & 33.8 \\ 
II & 37.7 & 34.1 & 28.2 \\ 
III & 43.2 & 32.6 & 24.2 \\ 
\end{tabular}
\caption{Nominal target and the resulting film compositions. Film II was analyzed by inductively coupled plasma optical emission spectroscopy (ICP-OES) and energy dispersive x-ray spectroscopy (EDX). Identically prepared films from targets I and III were investigated by EDX and corrected using the ICP-OES values of film II. The observed edge jump heights in the x-ray absorption spectra of Co and Mn agree with these concentrations.}
\label{tab:compositions}
\end{table}
The nominal target and the resulting film compositions are listed in Tab. \ref{tab:compositions}.
The films were capped by a thin Al layer to prevent unintentional oxidation of the Co-Mn-Sb or---in combination with plasma oxidation---to form a tunnel barrier. The layer sequences and notation of the samples discussed in this paper are given in Tab. \ref{tab1}.
\begin{table*}[htb]
\begin{tabular}{lllll}
{\bf sample}&{\bf layer sequence} & & & \\ \hline 
Half A$^{I,II,III}$& Si$_3$N$_4$-membrane/ & V 20\,(nm)/ & Co-Mn-Sb 40/& Al 1.6\\ 
Half B$^{I,II,III}$& Si-wafer / SiO$_2$ 50/ & V 40/ &  Co-Mn-Sb 100/& Al 1.6\\  
Half C$^{II}$& Si-wafer / SiO$_2$ 50/ & V 40/ &  Co-Mn-Sb 100/& Al 0-3 nm wedge + 150\,s plasma oxidation\\ 
Full$^{II}$& Si-wafer / SiO$_2$ 50/ & V 40/ & Co-Mn-Sb 100/& Al 0-3 nm wedge + 150\,s plasma oxidation\\
& + 60\,min 350$^\circ$C {\it in-situ} & annealing & + 50\,s plasma & oxidation + Co$_{70}$Fe$_{30}$ 5/ Mn$_{83}$Ir$_{17}$ 10/ + capping\\ 
\end{tabular}
\caption{Layer sequences of all samples. The sample series Half~A and Half~C were prepared for XAS and XMCD studies.  Structural and magnetic bulk properties were determined on Half~B samples. The transport properties of MTJs were measured on series Full. The roman superscript indicates the target number.}
\label{tab1}
\end{table*}

\section{Thin {Co}-{Mn}-{Sb} films on V buffers: Bulk properties}\label{SecIII}
\subsection{Magnetization studies: Alternating gradient field magnetometry (AGM)}\label{subsecA}
\begin{figure}
\begin{center}
\includegraphics[width=8cm]{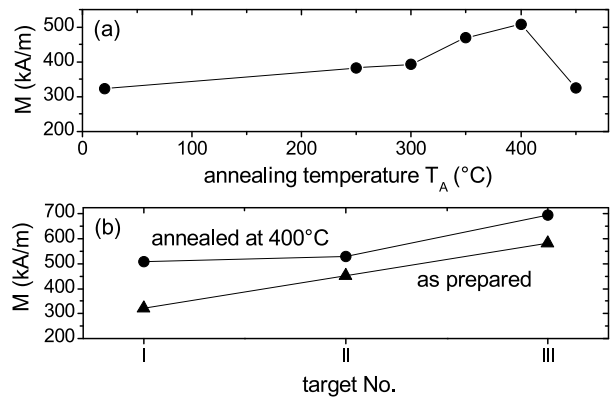}
\caption{Bulk saturation magnetization $M_S$ at room temperature: (a) Annealing temperature $T_A$ dependence of $M_S$ for sample Half~B$^I$. (b) Influence of target composition on $M_S$ for samples of type Half~B. }
\label{fig:1}
\end{center}
\end{figure}

The bulk magnetization of the samples was measured at room temperature in an AGM by Princeton Measurements, Inc. The saturation value $M_S$ of sample Half~B$^I$ in dependence of the annealing temperature $T_A$ is shown in Figure \ref{fig:1}a.
The largest RT magnetization of 508\,kA/m is found after annealing at 400$^\circ$C. 
The magnetization at low temperature is significantly higher than at RT because of the relatively low Curie temperature of about 480\,K\cite{haa87,fel06}. We observed a $15\%$ reduction of the Mn spin moment of sample Half~A$^I$ between $T=15$ K and RT (cp. Sec. \ref{subsecC}). Therefore we estimate $M_S$(15 K)$=600$\,kA/m. 
A similar increase of the magnetization with increasing annealing temperature is found for sample Half~B$^{II,III}$ (see Fig. \ref{fig:1}b). Furthermore, the bulk magnetization increases with increasing Co concentration of the films, which is highest for target III. 
\subsection{Inter-diffusion of Co-Mn-Sb and V: Auger depth profiling}\label{subsecAES}
\begin{figure}
\begin{center}
\includegraphics[width=8cm]{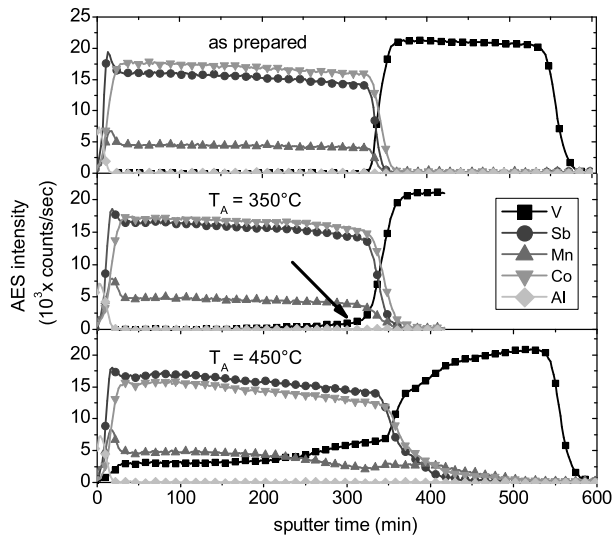}
\caption{Auger electron spectroscopy (AES) depth profiles of sample Half~B$^I$. V diffuses into Co-Mn-Sb at high annealing temperature. The depth resolution is not influenced by the annealing procedure\cite{DRF}.}
\label{fig:TP}
\end{center}
\end{figure}

The decrease of the magnetic moment above the optimal annealing temperature is caused by inter-diffusion of the Co-Mn-Sb layer with the V buffer. This is identified by Auger depth profiling\cite{wec02a,Sea90c} of sample Half~B$^{I}$ samples in the as prepared state and after annealing at 350$^\circ$C and 450$^\circ$C  (Figure \ref{fig:TP}). The different layers are well separated in the as prepared state. The apparent maximum of the Mn intensity just below the barrier is caused by preferential sputtering of Manganese. It is a very common artifact of sputter depth profiling\cite{Sea90c} and was also found for Antimony. After annealing at 350$^\circ$C the depth profile is slightly changed, a small fraction of Vanadium could be detected in the lower part of the Co-Mn-Sb layer (see arrow in Figure \ref{fig:TP}). This inter-diffusion intensifies with increasing annealing temperature. Therefore the annealing temperature for samples Half~A, Half~C$^{II}$ and Full$^{II}$ was generally set to 350$^\circ$C to minimize diffusion. 

In addition Auger depth profiles were taken for the Co rich samples Half~B$^{III}$. After annealing  at 450$^\circ$C the V concentration in the depth region of the Co-Mn-Sb layer was a factor of two smaller for sample Half~B$^{III}$ compared to sample Half~B$^{I}$, i.e. the inter-diffusion at the Co-Mn-Sb / V interface is considerably smaller for the off-stoichiometric samples. This indicates that the number of vacancy sites via which bulk diffusion can easily occur in Co-Mn-Sb decreases with increasing Co concentration. 
\subsection{Element-specific magnetic moments: X-ray absorption spectroscopy (XAS) and magnetic circular dichroism (XMCD)}\label{subsecC}
Temperature dependent x-ray absorption spectroscopy (XAS) and x-ray magnetic circular dichroism (XMCD) was performed at BL 6.3.1 of the Advanced Light Source in Berkeley, CA. The element-specific magnetic bulk properties were investigated at the Co- and Mn-$L$ edges in bulk-sensitive transmission mode\cite{kao94} on Half~A samples in the as-prepared state and after annealing at 350$^\circ$C. The x-rays angle of incidence with respect to the sample surface was $\alpha$=45$^\circ$. The XMCD spectra were obtained by applying a magnetic field of $\pm$ 0.2 T along the x-ray beam direction using elliptically polarized radiation with a polarization of $P_{h\nu}$=+75\%. $\mu^+$ and $\mu^-$ denote the linear absorption coefficients for parallel and anti-parallel orientation of photon spin and magnetic field. The transmitted x-rays were detected by a photo-diode. 
In conventional sum rule analysis\cite{set95} the number of holes $N_h$ needs to be taken from band structure calculations to calculate element-specific spin and orbital magnetic moments. Here, we follow the approach by St\"ohr {\it et al.}\cite{sto95} which bases on the assumption that the transition matrix elements connecting $2p$ core and $3d$ valence states are known and can be taken from other experiments. Details of our sum rule analysis procedure are given elsewhere \cite{rei09a}.

\begin{figure}
\begin{center}
\includegraphics[width=8cm]{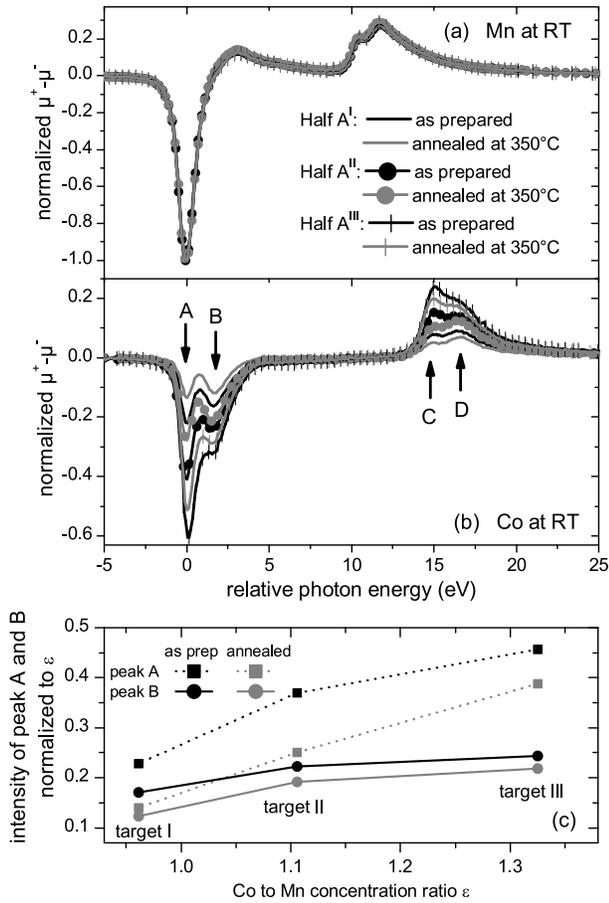}
\caption{XMCD spectra for sample Half~A at the Mn-$L_{2,3}$ (a) and Co-$L_{2,3}$ (b) edge. The spectra are normalized to the maximal XMCD signal at the Mn-$L_3$ edge for each sample. (c) Intensity of peak A and B normalized to the Co to Mn concentration ratio $\epsilon$.}
\label{fig:3}
\end{center}
\end{figure}

The XMCD spectra are shown in Fig. \ref{fig:3}a and b, the extracted magnetic moment ratios are summarized in Tab. \ref{tab:3}. 
\begin{table}[b]
\begin{tabular}{ccccc}
{\bf target}&  & $m_{orb}^{Mn}/m_{spin}^{Mn}$& $m_{orb}^{Co}/m_{spin}^{Co}$  &  $m^{Co}_{spin}/m^{Mn}_{spin}$\\ \hline 
I&as prep. & $-5.9\%$ & $7.6\%$ & $27.4\%$ \\
& 350$^\circ$C& $-5.3\%$ &  $0.3\%$  & $ 18.8\%$ \\ \hline
II&as prep. &  $-6.4\%$& $7.4\%$  & $ 38.5\%$ \\
& 350$^\circ$C&$-7.3\%$  &  $5.3\%$  & $ 29.6\%$ \\ \hline
III&as prep. & $-5.5\%$ & $7.0\%$ & $ 45.5\%$ \\
& 350$^\circ$C& $-5.5\%$ &  $6.1\%$ & $ 38.5 \%$ \\ \hline
\end{tabular}
\caption{Results of the sum rule analysis of the XMCD spectra shown in Fig. \ref{fig:3}a and b for sample Half~A.}
\label{tab:3}
\end{table}
The shape of the Mn XMCD asymmetry is nearly the same for all samples and very similar to Mn in the full Heusler compound Co$_2$MnSi\cite{sch04}. Accordingly, the Mn orbital to spin moment ratio does not depend on layer composition or post-annealing. The small Mn orbital moments are typically -6\% of the spin moments, i.e. both are aligned anti-parallel. Furthermore, the Mn magnetic moment of sample Half~A$^{I}$ was found to be reduced by $15\%$ between 15\,K and RT, which is in fair agreement with the data by Otto {\it et al.}\cite{haa87}, who observed a reduction of the saturation magnetization for bulk samples of $20\%$ between $T=4$\,K and RT and a Curie temperature of 478\,K.

The Co spin moment is aligned in parallel to its small orbital moment and the target composition and the post-annealing has a significant influence on the Co to Mn spin magnetic moments ratio which can be directly seen from the different sizes of the normalized Co XMCD signals in Fig. \ref{fig:3} b. Two clear trends are observed: First, the normalized Co XMCD signal and, therefore, the relative contribution of each Co atom to the total magnetic moment of the samples is reduced by annealing  for all samples of type Half~A.  Second, the relative contribution of the Co magnetic moments to the total magnetic moment of the samples increases with increasing Co concentration in the films. The smallest spin magnetic moment ratio $m^{Co}_{spin}/m^{Mn}_{spin}$ of 18.8\% is found for the annealed sample deposited from target I  while the largest one (45.5\%) is found for the as prepared sample deposited from target III (see Tab. \ref{tab:3}). Similar results were found for annealed Co-Mn-Sb films grown on Cu/Ta/V buffers and are published elsewhere\cite{rei09a}.

Moreover, there is a remarkable double peak structure of the Co XMCD spectra at the $L_2$ and $L_3$ edge, which is significantly different from Co metal\cite{set95} or Co$_2$MnSi\cite{sch04} and leads to a strongly changed electronic structure of the unoccupied $3d$ Co states compared to the references. This structure changes with different film composition and post-annealing. In addition to the decrease of the Co to Mn spin moment ratio, which corresponds to generally reduced intensities of the peaks "A" and "B" (Fig. \ref{fig:3}c), the intensity of peak "A" is reduced further in comparison to  "B" with decreasing Co concentration and by annealing. The same holds for the corresponding peaks "C" and  "D" at the  $L_2$ edge. 
The changed shape of the Co XMCD spectra could be caused by a different degree of location of the 3d states\cite{miy08} or hint to two non-equivalent Co positions  in the Co-Mn-Sb crystal lattice. This is in accordance with the structural model of alternating Co$_2$MnSb and MnSb structural units as discussed below. Moreover, because of the strong influence of the Co concentration on the spectral shape it is likely to assume, that the additional Co occupies the vacancy sites in the MnSb structural units.

The generally observed ferromagnetic coupling of the Mn and Co spin moment is important with respect to the different crystal structures discussed in the literature (see Sec. \ref{intro}). In the case of a C1$_b$ lattice an anti-ferromagnetic coupling of Co and Mn spin moments is expected ($m^{Co}_{spin}/m^{Mn}_{spin} = -4.1\%$). For the more complex structures resulting in a very narrow or even vanishing spin-down gap, both moments are expected to be ferromagnetically coupled (Senateur's model: $m^{Co}_{spin}/m^{Mn}_{spin} =+13.7\%$\cite{pie00}; disordered $L2_1$ structure: $m^{Co}_{spin}/m^{Mn}_{spin} =8.3\%$\cite{sko99}; alternation of Co$_2$MnSb and MnSb structural units: $m^{Co}_{spin}/m^{Mn}_{spin} =+15.8\%$\cite{fel06}). Further, for CoMnSb crystallized in the C1$_b$ structure a parallel alignment of the spin and orbital moments is expected\cite{gal05}. Obviously, our experimental data for the annealed Half A$^{I}$ sample fits best together with the structural model proposed by Ksenofontov {\it et al.}\cite{fel06}. 

By comparing the element-specific results and the bulk magnetization discussed in Sec. \ref{subsecA} one can conclude that the general increase of the magnetization with increasing Co content (see Fig. \ref{fig:1}) results from the larger Co content {\it and} the larger contribution of each Co atom to the total magnetic moment. The general reduction of $m^{Co}_{spin}/m^{Mn}_{spin}$ after annealing and the simultaneous increase of the bulk magnetization means, that the atomic order is improved by the post-annealing process: the magnetization of CoMnSb crystallized in one of the three complex structures suggested by Tobola {\it et al.}\cite{pie00}, Kaczmarska {\it et al.}\cite{sko99} and Ksenofontov {\it et al.}\cite{fel06}, respectively, is dominated by the Mn spin moment, which is aligned in parallel to the Co spin moment. Thereby, the Co magnetic moment is considerably smaller in these compounds compared to pure Co metal ($m^{Co}_{spin}\approx 1.6\mu_B$\cite{set95}), where nearest-neighbor Co atoms couple ferromagnetically. From this one would expect an increase of the mean Co magnetic moment per atom in case of atomic disorder, e.g., when Co atoms occupy Mn lattice sites. On the other hand, nearest-neighbor Mn atoms tend to couple antiferromagnetically, accordingly a decrease of the mean Mn magnetic moment would be expected for a disordered compound. This behavior has been predicted by Picozzi {\it et al.} for defects in $L2_1$ ordered Co$_2$MnSi\cite{fre04}. Accordingly, the results can be explained as an improvement of the atomic order by the post-annealing process. A similar behavior has been observed for full Heusler compound thin films and multilayers like Co$_2$MnSi\cite{sch04}, Co$_2$FeSi and $\{$Co$_2$MnSi / Co$_2$FeSi$\}_{\times 10}$\cite{rei06,are07b}.
\subsection{Crystal structure: X-ray diffraction (XRD)}\label{subsecD}
\begin{figure}
\begin{center}
\includegraphics[width=6cm]{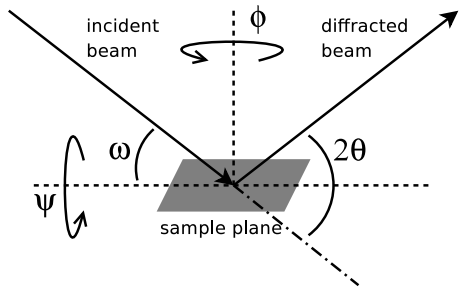}
\caption{The scattering geometry and the axes of the Philips X'Pert Pro MPD with open Eulerian cradle.}
\label{fig:xrd-angles}
\end{center}
\end{figure}
A Phillips X'Pert Pro MPD diffractometer equipped with an open Eulerian cradle and a copper anode was used for non-specular x-ray diffraction. Figure \ref{fig:xrd-angles} displays the scattering geometry and the angle definitions. Crystallinity and atomic order of Co-Mn-Sb was studied and related to the magnetic and electronic properties. 

Rocking curves and pole figures were taken to characterize the texture of the V buffer and the Co-Mn-Sb. Both materials grow (110) textured. The full width at half maximum (FWHM) of the Co-Mn-Sb (440)/(220) rocking curves is 8$^\circ\omega$ for sample Half~B$^I$ and 10$^\circ\omega$ for sample Half~B$^{III}$. 

In textured samples, only one group of diffraction peaks is accessible with $\theta$-$2\theta$ scans. One can find that the (220) peak intensity of Heusler compounds is independent of site-swap disorder\cite{AH25} by an analysis of the structure factors of a C1$_b$ (half) or L2$_1$ (full-)Heusler compound. In contrast, the (200) and (111) peak intensities depend on disorder and will vanish for A1 type disorder, where the lattice sites are randomly occupied. Therefore a (110)-textured L2$_1$ film will give the same specular diffractogram as a (110)-textured A1-disordered film of the same compound. Furthermore, this is also true for the various crystal structures which have been proposed for CoMnSb (see Sec. \ref{intro}).

\begin{figure}
\begin{center}
\includegraphics[width=8cm]{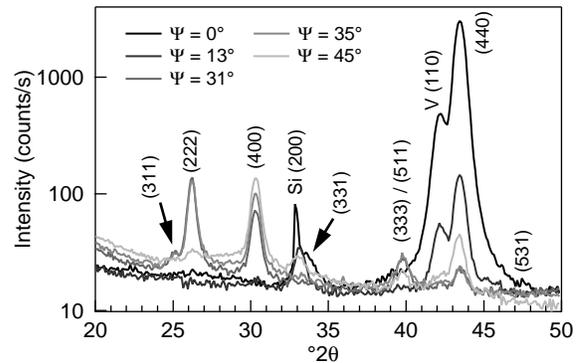}
\caption{Diffraction patterns of sample Half~B$^I$ (400$^\circ$C) taken under various tilt angles $\Psi$. The peaks not labeled by V or Si correspond to Co-Mn-Sb.}
\label{fig:FigXRDPsi}
\end{center}
\end{figure}

To overcome this problem, non-specular $\omega$-$2\theta$ scans were taken with various settings of $\psi$ and $2\theta = 2 \cdot \omega$. Figure \ref{fig:FigXRDPsi} presents a set of these scans taken on sample Half~B$^I$ (400$^\circ$C). The scan with $\psi = 0^\circ$ shows the V (110) reflex and the strong (440) reflex from the Co-Mn-Sb. At higher values of $\psi$ more reflections appear. A well-ordered C1$_b$ lattice would exhibit the (222), (400) and (440) reflections only (which would be indexed (111), (200) and (220) then). The additional (311), (331), (333)/(511) and (531) peaks have comparatively low intensity and grow with increasing annealing temperature. Similar patterns were found by Ksenofontov {\it et al.}\cite{fel06} in powders of Co-Mn-Sb. The peaks can not be indexed based on a L2$_1$/C1$_b$ structure with a lattice constant of about 5.9\,\AA{}. In contrast, they can be indexed by the assumption of a doubled unit cell parameter of 11.77\,\AA{}. 

Analogous measurements on sample Half~B$^{III}$ did not show the additional peaks. The complex unit cell with a doubled lattice constant is obviously specific for the nearly stoichiometric CoMnSb. The Co excess in films sputtered from target III might lead to occupation of the vacant sites in the MnSb substructure of the Co$_2$MnSb-MnSb super cells. If finally one Co atom per f.u. is added (without reducing the Sb concentration at the same time), the system will be in the $L2_1$ structure.   

\begin{figure}
\begin{center}
\includegraphics[width=8cm]{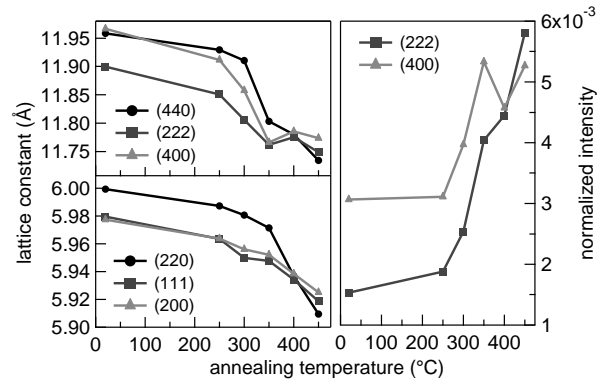}
\caption{Left: Lattice constants of sample Half~B$^I$ (top) and sample Half~B$^{III}$ (bottom), determined on three independent reflexes at the corresponding tilt angles. Right: Normalized integrated intensity of the (222) and (400) peaks for sample Half~B$^I$. (220)/(440) reflections were taken at $\Psi = 0.00^\circ$, (111)/(222) at $\Psi = 35.26^\circ$ and (200)/(400) at $\Psi = 45.00^\circ$.}
\label{fig:peakanalysis}
\end{center}
\end{figure}

In Figure \ref{fig:peakanalysis} we determine the in-plane and out-of-plane lattice parameters as a function of the annealing temperature for sample Half~B$^I$ and sample Half~B$^{III}$. The lattice parameters shrink with increasing annealing temperature, indicating a lattice relaxation and/or ordering. After 400$^\circ$C the lattice parameters in the three probed directions are the same, stress-induced influences might be compensated.

The integrated intensity of the (222) and (400) reflexes is a measure for the order in the films. Figure \ref{fig:peakanalysis} (right) shows a plot of the normalized integrated (222) and (400) peak intensities as a function of the annealing temperature. The normalization is done with respect to the integrated (440) intensity to correct for recrystallization. The (222) intensity increases by a factor of 4 from the as-prepared sample to 450$^\circ$C. Note that the intensity of the (222) reflex is further increased with annealing from 400$^\circ$C to 450$^\circ$C, whereas the (400) reflex intensity does not increase monotonously, but rather shows a dip at $T_A$=400$^\circ$C. 

The additional peaks of the CoMnSb superstructure vanish for annealing at $T_A$=450$^\circ$C. Vanadium is implemented into the lattice and thus decreases the saturation magnetization (Section \ref{subsecA}). The dip in the normalized (400) intensity curve at 400$^\circ$C hints the incipient diffusion of V, changing the properties of the lattice. However, the face-centered cubic structure is preserved and the V increases the structure factor of the (222) reflex. 

\section{The Co-Mn-Sb / Al-O interface}\label{SecIV}
\subsection{X-ray absorption spectroscopy (XAS)}\label{subsecB}
\begin{figure}
\begin{center}
\includegraphics[width=8cm]{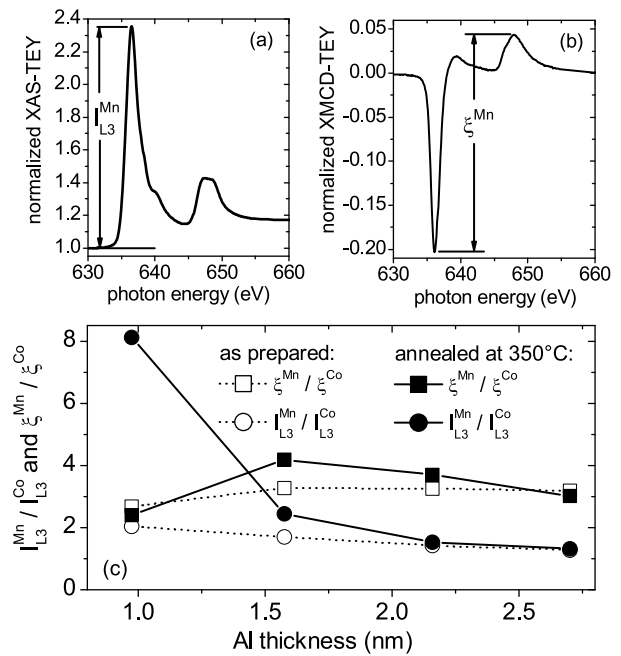}
\caption{(a) XA-TEY spectrum at the Mn $L_{2,3}$ edges of  sample Half~C$^{II}$ (Al thickness: 1.6\,nm) after {\it ex-situ} annealing at 350$^\circ$C. The maximum intensity of the $L_3$ resonance is defined as $I_{L3}^{Mn}$. (b) Corresponding XMCD-TEY spectrum.  The definition of the maximum XMCD asymmetry $\xi^{Mn}$ is shown. (c) Summary of the intensity and asymmetry ratios  $I_{L3}^{Mn} / I_{L3}^{Co}$ and $\xi^{Mn} / \xi^{Co}$ of sample Half~C$^{II}$. }
\label{fig:5x}
\end{center}
\end{figure}
\begin{figure}
\begin{center}
\includegraphics[width=8cm]{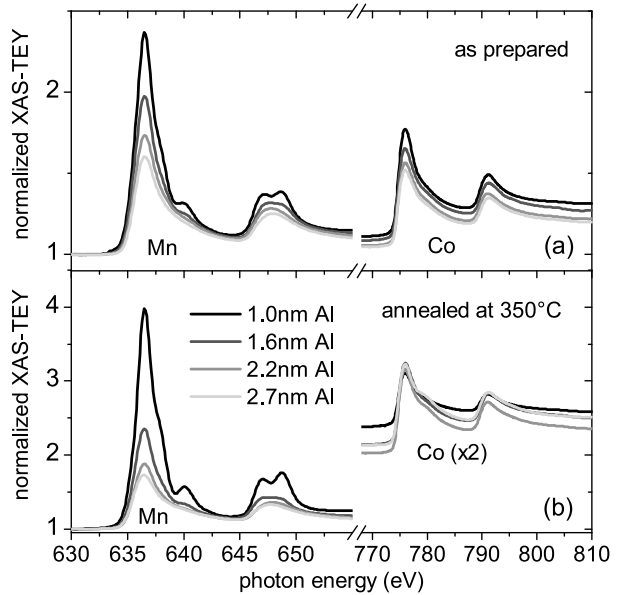}
\caption{XA-TEY spectra at the Co and Mn $L_{2,3}$ edges of sample Half~C$^{II}$.}
\label{fig:5y}
\end{center}
\end{figure}
The influence of the Al thickness and the annealing on the chemical states of Co and Mn at the Co-Mn-Sb / Al-O interface was investigated by surface-sensitive XAS and XMCD in total electron yield (TEY) mode applied to Half~C$^{II}$ samples, involving a plasma oxidized wedge of Al on top of Co-Mn-Sb. The Co-Mn-Sb/Al-O interface needs to be optimized with respect to the oxidation process of the Aluminum in order to reach maximum TMR ratios \cite{rei04c,sac08,are07b}.

An increasing Mn to Co concentration ratio was found at the interface with decreasing Al thickness (see $I_{L3}^{Mn} / I_{L3}^{Co}$ in Fig. \ref{fig:5x}c). This trend becomes more pronounced after annealing. The shapes of the Mn XA spectra in Figure \ref{fig:5y} show that this Mn segregation at the Co-Mn-Sb/Al-O interface is caused by MnO formation at the lower barrier interface. For an Al thickness of only 1.0\,nm the spectra of the as prepared as well as the annealed samples were dominated by the multiplet structure typical for MnO\cite{oht01}. The multiplet structure became weaker with increasing Al thickness. A structureless $L_{2,3}$ absorption edge was found for thick Aluminum, its shape is similar to metallic bulk Mn\cite{oht01}. 

The significant formation of interfacial MnO (paramagnetic at room temperature\cite{pep74}) for 1.0\,nm thick Al is also reflected in the low XMCD asymmetry ratio $\xi^{Mn} / \xi^{Co}$. This is proportional to the ratio of the Mn and Co magnetic moments, which are both dominated by the spin moment (see above). $\xi^{Mn} / \xi^{Co}$ was maximal after annealing for an Al thickness of 1.6\,nm and the contribution of MnO to the XA spectrum is already very small compared to 1.0\,nm Al (see Figure \ref{fig:5y}b). Here, the best atomic order was achieved, since disorder is expected to {\it reduce} the Mn to Co magnetic moment ratio (see Sec. \ref{subsecC}). While thin Al leads to an oxidation of the ferromagnetic electrode, metallic Al will remain at the interface, if the Al thickness is too large. This Al can diffuse into the Co-Mn-Sb electrode during annealing leading to a reduction of the interfacial magnetic moments and the Mn to Co magnetic moment ratio. The observed increase of the Mn to Co (bulk) spin moment ratio of all films discussed in the previous Sec. \ref{subsecC} was not found at the Co-Mn-Sb/Al-O interface  for 2.7 nm thick Al due to residual unoxidized Al. However, such an increase was found for optimum aluminum thickness of 1.6\,nm. At this Al thickness, the properties of the Co-Mn-Sb near the surface are closest to those of the bulk material. 
\subsection{Transport properties of magnetic tunnel junctions with Co-Mn-Sb electrode}\label{subsecE}
The magnetic tunnel junctions Full$^{II}$ were prepared with optimized growth conditions as discussed earlier: they were {\it in-situ} annealed at 350$^\circ$C after 150\,s plasma-oxidation of the Al wedge. In order to remove contaminations from the surface, the samples were plasma oxidized for another 50\,s after the annealing. Subsequently, {\it ex-situ} vacuum annealing at 275$^\circ$C in a magnetic field of 0.1\,T was employed to set the exchange bias of the pinned electrode. In previous papers\cite{kam04,rei06} we show that this preparation scheme is suitable to produce magnetic tunnel junctions with Heusler compound electrodes that exhibit high TMR ratios. Patterning was done by optical lithography and ion beam etching, the transport properties of the quadratic 100 $\times$ 100 $\mu$m$^2$ wide MTJs were measured by conventional two-probe method with a constant dc bias voltage. 

\begin{figure}
\begin{center}
\includegraphics[width=8cm]{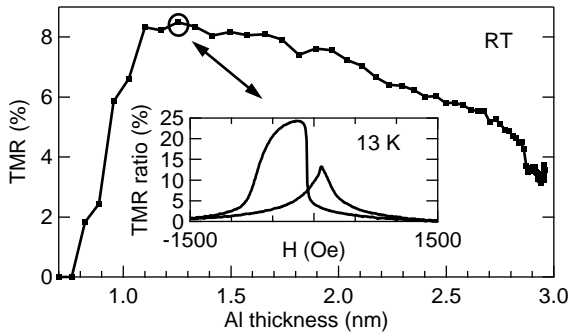}
\caption{TMR as a function of the Al thickness of Full$^{II}$. The measurements were taken with a dc bias voltage of 10\,mV.}
\label{fig:wedge}
\end{center}
\end{figure}

In Figure \ref{fig:wedge} the dependence of the TMR on the Al thickness is shown. A maximal TMR ratio of 8.5\% at room temperature was found at an Al thickness of 1.25 nm (1.6 nm AlOx). The Al thickness dependence of $\xi^{Mn} / \xi^{Co}$ presented in the last section fits quite well to the Al thickness dependence of the TMR. The observed maximum TMR ratio at 13\,K was 24\% (see inset in Figure \ref{fig:wedge}). According to the Julli\`ere model, the effective tunneling spin polarization is calculated to be 0.22, using a spin polarization for Co$_{70}$Fe$_{30}$ of 0.5 \cite{ThomasPhD}. No higher values for the TMR ratio could be found with varied film growth parameters. Moreover, MTJs with Co-Mn-Sb electrodes from targets I and III show comparable TMR ratios. Taking this together with the careful preparation of the films and the tunnel barrier, we conclude that the low observed spin polarization is an intrinsic property of our Co-Mn-Sb electrodes. 
\section{Conclusion}
It was shown that post annealing increases the atomic order and the magnetization of the Co-Mn-Sb films, while the Co to Mn magnetic moment ratio decreased. The samples degraded for temperatures higher than 350$^\circ$C  because of inter-diffusion of V at the Co-Mn-Sb / V interface. Adding Co to the samples leads to higher magnetization values and to a larger Co to Mn magnetic moment ratio as expected from simple arguments basing on the typical nearest-neighbor coupling behavior of Co-Co and Mn-Mn pairs. 
A parallel coupling of the Co and Mn magnetic spin moments was found for all compositions. For nearly stoichiometric films the Co to Mn spin moment ratio was +18.8\% after annealing and the unit cell parameter was 11.77\,\AA{}. This fits the expectations for CoMnSb crystallized in a complex Co$_2$MnSb/MnSb super structure as reported by Ksenofontov {\it et al.}\cite{fel06}. The addition of Co as well as interdiffusion from the vanadium buffer destroy the super structure. 
Finally, the transport properties of magnetic tunnel junctions with Co-Mn-Sb electrode, Al-O barrier and Co-Fe counter electrode are in agreement with the predicted low spin polarization: we found an effective tunneling spin polarization of only 22\%, which clearly indicates a non-half metallic character of CoMnSb.

\section*{Acknowledgement}
The authors gratefully acknowledge financial support by the Deutsche Forschungsgemeinschaft (DFG) and the Deutscher Akademischer Austausch Dienst (DAAD) and the opportunity to work at BL 6.3.1 of the Advanced Light Source, Berkeley, USA. The ALS is supported by the U.S.\ Department of Energy under Contract No.\ DE-AC02-05CH11231. For financial support of the XRD measurements under contract No.DAAD/10/2006 M.M., T.S. and J.K. gratefully acknowledge the Ministry of Science and Higher Education of Poland.


\begin{thebibliography}{43}
\bibitem[{DRF()}]{DRF}
\bibinfo{note}{The depth resolution of the profiles can be calculated from the
  slope of the V intensity at the V / SiO$_2$ interface\cite{Sea90c}. The slope
  is very similar for the as prepared and the 450$^\circ$C samples, the sputter
  time needed to reduce the intensity from 84\% to 16\% of the maximum V
  intensity is 20.7min (as prepared) and 20.4min (450$^\circ$C), respectively.
  By comparing these sputter times with the required 214min to sputter through
  the 40nm thick V layer of the as prepared sample an identical depth
  resolution of about 4nm can be estimated for all three samples.}
\end{thebibliography}
\end{document}